\documentclass[preprint,prd,aps,showpacs,showkeys,nofootinbib]{revtex4}
\usepackage{graphicx}

\begin{document}

\title{Constraints on parameter space of BLMSSM from particle mass}

\author{Hui Li$^{a}$, Jian-Bin Chen$^a$\footnote{email:chenjianbin@tyut.edu.cn}, Li-Li Xing$^a$}

\affiliation{$^a$College of Physics and Optoelectronic Engineering, Taiyuan University of Technology, Taiyuan 030024, China}

\begin{abstract}
To explaine the matter-antimatter asymmetry,
a supersymmetric extention of the standard model is proposed where baryon and lepton numbers are local gauged(BLMSSM),
and exotic superfields are introduced when gauge group is enlarged to $SU(3)_C\otimes SU(2)_L\otimes U(1)_Y \otimes U(1)_B \otimes U(1)_L$.
As signals of new physics have not been observed on Large Hadron Collider,
the parameter space relevant to the masses of new particles is stringently constrainted.
By diagonalizing the mass squared matrices for neutral scalar sectors and the mass matrices for exotic quarks,
we plot the masses of new particles varying with different parameters with some assumptions,
so the constraints on model parameter is obtained with different lower limit on particle mass.
\end{abstract}


\maketitle

\section{Introduction\label{sec1}}
As an effective theory, the Standard Model(SM) is considered to be the most successful theory of particle physics. Though the discovery of Higgs boson makes SM seems complete\cite{bibitem1,bibitem2}, it dose not provide candidate of dark mater. The explanation to hierarchy problem is another motivation for people to propose new physics beyond SM. By introducing the symmetry between fermions and bosons, supersymmetry(SUSY) solves the hierarchy problem naturally,
and the Minimal Supersymmetric Standard Model(MSSM) is the simplest one\cite{bibitem3,bibitem4,bibitem5,bibitem6}.
With R-parity preserved, the lightest superparticle (LSP) of the MSSM is a good dark matter candidate.
Since the asymmetry of matter-antimatter causes baryon number break, an extension with baryon and lepton numbers local gauged is proposed with enlarged gauge group $SU{(3)_C} \otimes SU{(2)_L} \otimes U{(1)_Y} \otimes U{(1)_B} \otimes U{(1)_L}$.

In this model, the baryon and the lepton numbers are local gauge symmetries spontaneously broken at the TeV scale\cite{bibitem7,bibitem8,bibitem9}, there is not dangerous baryon number violating operators. The generation of the heavy majorana neutrinos can understand the tiny mass of the neutrinos by the seesaw mechanism\cite{bibitem10,bibitem11,bibitem12,bibitem13,bibitem14,bibitem15,bibitem16,bibitem17}. So BLMSSM is R-parity violating and can suppress the flavor violation in the quark and leptonic sectors\cite{bibitem18,bibitem19,bibitem20,bibitem21}.
One adds new quarks with B=1, and new leptons with L=3, to cancel fermionic family anomalies. In addition, the model ignore the coupling constants with Landau poles near the weak scale and there is no flavor changing neutral currents at tree level, and hence one does not need ``desert region''.
Futhermore, the model forbid proton decay\cite{bibitem22} and ensure the stable dark matter candidate.

In this work, we study the masses of new particles in the framework of BLMSSM. As the gauge group is enlarged, many new parameters are introduced in this model. By extracting the mass matrices from superpotential and soft breaking terms, the masses can be expressed with model parameters.
Therefor, it's necessary to constrain the parameters space with the lower limits suggested by LHC obsearvation.
This paper has following structure. The BLMSSM model is introduced in Section 2, In section 3, we summarize the mass matrices. Section 4 shows the numerical analysis and discussions of parameter space. In section 5, the conclusion is given.

\section{The supersymmtric extension of BLMSSM\label{sec2}}

 When two new gauge symmetries are introduced, mean that baryon and lepton number are gauged, hence the local gauge group is extended to $SU{(3)_C} \otimes SU{(2)_L} \otimes U{(1)_Y} \otimes U{(1)_B} \otimes U{(1)_L}$. This model is called BLMSSM.
 New fields in the BLMSSM are collected here: the new quarks ${{\hat Q}_4} \sim (3, 2, 1/6, {B_4}, 0), $ $\hat U_4^C \sim (\overline 3, 1, - 2/3, - {B_4}, 0),$ $\hat D_4^C \sim (\overline 3, 1, 1/3, - {B_4}, 0),$
$\hat Q_5^C \sim (\overline 3, 2, - 1/6, -(1 + {B_4}), 0),$  ${{\hat U}_5} \sim (3, 1, 2/3, 1 + {B_4}, 0),$ ${{\hat D}_5} \sim (3, 1, - 1/3, 1 + {B_4}, 0)$ , and the new leptons ${{\hat L}_4} \sim (1, 2, - 1/2, 0, {L_4}),$ $\hat E_4^C \sim (1, 1, 1, 0, - {L_4}),$
$\hat N_4^C \sim (1, 1, 0, 0, - {L_4}),$
$\hat L_5^C \sim (1, 2, 1/2, 0, - (3 + {L_4})),$ ${{\hat E}_5} \sim (1, 1, - 1, 0, 3 + {L_4}),$ ${{\hat N}_5} \sim (1, 1, 0, 0, 3 + {L_4})$,
to cancel the B and L anomalies, respectively.
To avoid having a stable exotic quarks the model introduce the superfields $\hat X \sim (1, 1, 0, 2/3 + {B_4}, 0)$,
and ${\hat X^{'}} \sim (1, 1, 0, - (2/3 + {B_4}), 0)$\cite{bibitem23,bibitem24,bibitem25}.
Meanwhile, to break L and B spontaneously with nonzero vacuum expectation values (VEVs) and  to provide masses for the exotic leptons and exotic quarks,
the BLMSSM add the exotic Higgs superfields ${\hat \phi _L}$, ${\hat \varphi _L}$ and ${\hat \phi _B}$, ${\hat \varphi _B}$.
In addition, using the right-handed neutrinos $N_R^C$, tiny masses related to neutrinos are acquired by see-saw mechanism.

The dark matter candidate corresponding to the lightest mass eigenstate, when $\hat X$ and ${\hat X^{'}}$ mix together.
In terms of the BLMSSM, the superpotential is written as\cite{bibitem26}
\begin{eqnarray}
{W_{BLMSSM}} = {W_{MSSM}} + {W_B} + {W_L} + {W_X},
\end{eqnarray}
with ${W_{MSSM}}$ denoting the superpotential of the MSSM.
The superpotential of the baryon, lepton are denoted respectively\cite{bibitem26}:
\begin{eqnarray}
{W_B}&=&{\lambda _Q}{\hat Q_4}\hat Q_5^C{\hat \phi _B} + {\lambda _U}\hat U_4^C{\hat U_5}{\hat \varphi _B} + {\lambda _D}\hat D_4^C{\hat D_5}{\hat \varphi _B} + {\mu _B}{\hat \phi _B}{\hat \varphi _B}\nonumber\\
&& + {Y_{{{\text{u}}_4}}}{\hat Q_4}{\hat H_u}\hat U_4^C + {Y_{{{\text{d}}_4}}}{\hat Q_4}{\hat H_d}\hat D_4^C + {Y_{{{\text{u}}_5}}}\hat Q_5^C{\hat H_d}{\hat U_5}
 + {Y_{{{\text{d}}_5}}}\hat Q_5^C{\hat H_u}{\hat D_5},\nonumber\\
{W_L} &=&{Y_{{e_4}}}{\hat L_4}{\hat H_d}\hat E_4^c + {Y_{{\nu _4}}}{\hat L_4}{\hat H_u}\hat N_4^c + {Y_{{e_5}}}\hat L_5^C{\hat H_u}{\hat E_5} + {Y_{{\nu _5}}}\hat L_5^c{\hat H_d}{\hat N_5}\nonumber\\
 &&+ {Y_\nu }\hat L{\hat H_u}{\hat N^c} + {\lambda _{{N^c}}}{\hat N^c}{\hat N^c}{\hat \varphi _L} + {\mu _L}{\hat \phi _L}{\hat \varphi _L},\nonumber\\
{W_X} &=& {\lambda _1}\hat Q\hat Q_5^c\hat X + {\lambda _2}{\hat U^c}{\hat U_5}{\hat X^{'}} + {\lambda _3}{\hat D^c}{\hat D_5}{\hat X^{'}} + {\mu _X}\hat X{\hat X^{'}}.
\end{eqnarray}
 Correspondingly, the soft breaking terms of the MSSM are given as\cite{bibitem26,bibitem27}
\begin{eqnarray}
\mathcal{L} = &{\mathcal{L}}_{soft}^{MSSM} - {(m_{{{\tilde \nu }^c}}^2)_{IJ}}\tilde N_I^{c*}\tilde N_J^c  - m_{{{\tilde Q}_4}}^2\tilde Q_4^\dag {{\tilde Q}_4} - m_{{{\tilde U}_4}}^2\tilde U_4^{c*}\tilde U_4^c\nonumber\\
& - m_{{{\tilde D}_4}}^2\tilde D_4^{C*}\tilde {D_4^c}  - m_{{{\tilde Q}_5}}^2\tilde Q_5^{c\dag }\tilde Q_5^c - m_{{{\tilde U}_5}}^2\tilde U_5^*{{\tilde U}_5} - m_{{{\tilde D}_5}}^2\tilde D_5^*{{\tilde D}_5}\nonumber\\
& - m_{{{\tilde L}_4}}^2\tilde L_4^\dag {{\tilde L}_4}- m_{{{\tilde \nu }_4}}^2\tilde N_4^{c*}\tilde N_4^c - m_{{{\tilde e}_4}}^2\tilde E_4^{c*}\tilde E_4^C - m_{{{\tilde L}_5}}^2\tilde L_5^{c\dag }\tilde L_5^c\nonumber\\
& - m_{{{\tilde \nu }_5}}^2\tilde N_5^*{{\tilde N}_5} - m_{{{\tilde e}_5}}^2\tilde E_5^*{{\tilde E}_5} - m_{{\phi _B}}^2\phi _B^*{\phi _B} - m_{{\varphi _B}}^2\varphi _B^*{\varphi _B}\nonumber\\
& - m_{{\phi _L}}^2\phi _L^*{\phi _L} - m_{{\varphi _L}}^2\varphi _L^*{\varphi _L} - ({m_B}{\lambda _B}{\lambda _B} + {m_L}{\lambda _L}{\lambda _L} + h.c.)\nonumber\\
&+ \{ {A_{{u_4}}}{Y_{{u_4}}}{{\tilde Q}_4}{H_u}\tilde U_4^c + {A_{{d_4}}}{Y_{{d_4}}}{{\tilde Q}_4}{H_d}\tilde D_4^c + {A_{{u_5}}}{Y_{{u_5}}}\tilde Q_5^c{H_d}{{\tilde U}_5}\nonumber\\
& + {A_{{d_5}}}{Y_{{d_5}}}\tilde Q_5^c{H_u}{{\tilde D}_5} + {A_{BQ}}{\lambda _Q}{{\tilde Q}_4}\tilde Q_5^c{\phi _B} + {A_{BU}}{\lambda _U}\tilde U_4^c{{\tilde U}_5}{\varphi _B}\nonumber\\
& + {A_{BD}}{\lambda _D}\tilde D_4^c{{\tilde D}_5}{\varphi _B} + {B_B}{\mu _B}{\phi _B}{\varphi _B} + h.c.\}\nonumber\\
& + \{ {A_{{e_4}}}{Y_{{e_4}}}{{\tilde L}_4}{H_d}\tilde E_4^c + {A_{{\nu _4}}}{Y_{{\nu _4}}}{{\tilde L}_4}{H_u}\tilde N_4^c + {A_{{e_5}}}{Y_{{e_5}}}\tilde L_5^c{H_u}{{\tilde E}_5}\nonumber\\
& + {A_{{\nu _5}}}{Y_{{\nu _5}}}\tilde L_5^c{H_d}{{\tilde N}_5} + {A_N}{Y_\nu }\tilde L{H_u}{{\tilde N}^c} + {A_{{N^c}}}{\lambda _{{N^c}}}{{\tilde N}^c}{{\tilde N}^c}{\varphi _L}\nonumber\\
& + {B_L}{\mu _L}{\phi _L}{\varphi _L} + h.c.\} + \{ {A_1}{\lambda _1}\tilde Q\tilde Q_5^cX + {A_2}{\lambda _2}{{\tilde U}^c}{{\tilde U}_5}X'\nonumber\\
& + {A_3}{\lambda _3}{{\tilde D}^c}{{\tilde D}_5}X' + {B_X}{\mu _X}XX' + h.c.\},
\end{eqnarray}
with $\lambda _B$ is gaugino of $U(1)_B$, $\lambda _L$ is gaugino of $U(1)_L$, respectively.

The local gauge symmetry $SU{(2)_L} \otimes U{(1)_Y} \otimes U{(1)_B} \otimes U{(1)_L}$ breaks down to the electromagnetic symmetry $U{(1)_e}$ if the following conditions are satisfied[30]:

 1. the Higgs field $SU{(2)_L}$ doublets $({H_u}, {H_d})$ obtain nonzero VEVs ${\nu _u}$, ${\nu _d}$.

  2. the Higgs field $SU{(2)_L}$ singlets $({\phi _L}, {\varphi _L}, {\phi _B},{\varphi _B})$ obtain nonzero VEVs ${\nu _B},{\bar \nu _B}, {\nu _L}, {\bar \nu _L}$.

Where the $SU{(2)_L}$ doublets $({H_u}, {H_d})$ and the $SU{(2)_L}$ singlets $({\phi _B}, {\varphi _B}, {\phi _L},{\varphi _L})$ are defined as
\begin{eqnarray}
  H_u &=&
  \left(
  \begin{array}{ll}
  H_u^{+} \\
  \frac{1}
{{\sqrt 2 }}({\nu _u} + H_u^0 + iP_u^0)
  \end{array}
  \right),\\
  H_d &=&
  \left(
  \begin{array}{ll}
\frac{1}
{{\sqrt 2 }}({\nu _d} + H_d^0 + iP_d^0)
 \\
H_d^{-}
  \end{array}
  \right),\\
{\phi _B} &=& \frac{1}
{{\sqrt 2 }}({\nu _B} + \phi _B^0 + iP_B^0),\\
{\varphi _B} &=& \frac{1}
{{\sqrt 2 }}({\bar \nu _B} + \varphi _B^0 + i\bar P_B^0),\\
{\phi _L} &=& \frac{1}
{{\sqrt 2 }}({\nu _L} + \phi _L^0 + iP_L^0),\\
{\varphi _L} &=& \frac{1}
{{\sqrt 2 }}({\bar \nu _L} + \varphi _L^0 + i\bar P_L^0).
\end{eqnarray}

Here, the values of VEVs ${\nu _u},{\nu _d},{\nu _B},{\bar \nu _B},{\nu _L},{\bar \nu _L}$ are nonzero, we adopt the shortcut notations $\tan\beta_B=\bar{v}_B/v_B$ and $\bar{v}_B^2+v_B^2=v_{bt}^2$.

\section{The mass matrices for some new particles\label{sec3} }
 From the soft breaking terms and superpotential of BLMSSM, we can extracted the mass matrices for following particles:

 1. Two charged Higgs scalars denoted by $H_1^{\pm}, (i=1, 2)$ introduced in MSSM are related to the initial Higgs fields by the rotation matrix $Z_H$ and that is defined as
\begin{eqnarray}
Z_H=
\left(\begin{array}{cc}\sin\beta  & -\cos\beta  \\
                       \cos\beta  &  \sin\beta  \end{array}\right) ,~~~~~
\text{with}~~~
\left(\begin{array}{c}H_2^{1*}  \\ H_1^2 \end{array}\right)
=Z_H
\left(\begin{array}{c}H_1^+ \\ H_2^+ \end{array}\right).
\end{eqnarray}

2. The masses of four neutralinos $\chi_i^0, (i=1, \cdots 4)$ can be obtained by diagonalizing the mass matrices with the rotation matrix $Z_{N}$ as follow
\begin{eqnarray}
Z_N^T
\left(\begin{array}{cccc}M_1 & 0 & \frac{-ev_1}{2c_W} & \frac{ ev_2}{2c_W} \\
                        0 & M_2  & \frac{ ev_1}{2s_W} & \frac{-ev_2}{2s_W} \\
                        \frac{-ev_1}{2c_W} & \frac{ ev_1}{2s_W} &0 & \mu    \\
                        \frac{ ev_2}{2c_W} & \frac{-ev_2}{2s_W} &-\mu & 0    \\
\end{array}\right)Z_N
=
\left(\begin{array}{ccc}M_{\chi_1^0} &  & 0 \\
                          & \ddots &  \\
                        0 & & M_{\chi_4^0}\\
\end{array}\right),
\end{eqnarray}
where $\chi_{i}^{0}, i=1\ldots4$ reprent four Majorana fermions.

3. Compared with MSSM, the baryon neutralinos $\chi_{B_j}^0$ are the exotic particles in the BLMSSM. The new gaugino $\lambda_B$ and the superpartner of the $SU(2)_{B}$ singlets $\phi_B$ and $\varphi_B$ mix, which produce three baryon neutralinos in the base (i$\lambda_B, \psi_{\phi_{B}}, \psi_{\varphi_{B}}$)\cite{bibitem28,bibitem29,bibitem30}.
\begin{eqnarray}
\frac{1}{2}
A\left(\begin{array}{ccc}2M_B & -v_Bg_B  & \bar{v}_Bg_B  \\
                        -v_Bg_B & 0  & -\mu_B  \\
                        \bar{v}_Bg_B & -\mu_B  & 0  \end{array}\right)
A^T,
\end{eqnarray}
with
\begin{eqnarray}
A =\left(\begin{array}{ccc}i\lambda_B, & \psi_{\Phi_B},  & \psi_{\varphi_B}\end{array}\right).
\end{eqnarray}

Using $Z_{N_{B}}$ , one can diagonalize the mass matrix in Eq. (14) to obtain three lepton neutralino masses .

4. In BLMSSM, the mass squared matrix of the Down-squarks is different from that in MSSM, because of two exotic part $\frac{g_B^2}{6}(v_B^2-\bar{v}^2_B)$ and $-\frac{g_B^2}{6}(v_B^2-\bar{v}^2_B)$ from $(\mathcal{M}^2_D)_{LL}$ and $(\mathcal{M}^2_D)_{RR}$.
\begin{eqnarray}
Z_D^\dag
\left(\begin{array}{cc}  (\mathcal{M}_D^2)_{LL}      & (\mathcal{M}_D^2)_{LR} \\
                         (\mathcal{M}_D^2)_{LR}^\dag & (\mathcal{M}_D^2)_{RR}
      \end{array}\right)Z_D
=
\left(\begin{array}{ccc}M_{D_1}^2 &  & 0 \\
                          & \ddots &  \\
                        0 & & M_{D_6}^2
\end{array}\right).
\end{eqnarray}

The unitary matrix $Z_D$ is used to rotate Down-squarks mass squared matrix to mass eigenstates. And $(\mathcal{M}^2_D)_{LL}$, $(\mathcal{M}^2_D)_{RR}$, $(\mathcal{M}^2_D)_{LR}$ are read as

\begin{eqnarray}
(\mathcal{M}^2_D)_{LL}&=&-\frac{e^2(v_1^2-v_2^2)(1+2c_W^2)}{24s_W^2c_W^2}+\frac{v_1^2Y_d^2}{2} +(m_Q^2)^T
+\frac{g_B^2}{6}(v_B^2-\bar{v}^2_B),\nonumber\\
(\mathcal{M}^2_D)_{RR}&=&-\frac{e^2(v_1^2-v_2^2)}{12c_W^2}+\frac{v_1^2Y_d^2}{2} +m_D^2-\frac{g_B^2}{6}(v_B^2-\bar{v}^2_B),\nonumber\\
(\mathcal{M}^2_D)_{LR}&=&\frac{1}{\sqrt{2}}\Big(v_2(-A_d'+Y_d\mu^*)+v_1A_d\Big).
\end{eqnarray}

we adopt the shortcut notations: $c_W = \cos\theta_{W}, s_W = \sin\theta_{W}$, with $\theta_{W}$ denoting the Weinberg angle.

5. In BLMSSM, the mass squared matrix of the Up-squarks is different from that in MSSM, because of two exotic part $\frac{g_B^2}{6}(v_B^2-\bar{v}^2_B)$ and $-\frac{g_B^2}{6}(v_B^2-\bar{v}^2_B)$ from $(\mathcal{M}^2_U)_{LL}$ and $(\mathcal{M}^2_U)_{RR}$,
\begin{eqnarray}
Z_U^T
\left(\begin{array}{cc}  (\mathcal{M}_U^2)_{LL}      & (\mathcal{M}_U^2)_{LR} \\
                         (\mathcal{M}_U^2)_{LR}^\dag & (\mathcal{M}_U^2)_{RR}
      \end{array}\right)
Z_U^*=
\left(\begin{array}{ccc}M_{U_1}^2 &  & 0 \\
                          & \ddots &  \\
                        0 & & M_{U_6}^2
\end{array}\right).
\end{eqnarray}
The unitary matrix $Z_U$ is used to rotate Up-squarks mass squared matrix to mass eigenstates.
 And $(\mathcal{M}^2_U)_{LL}$ ,$(\mathcal{M}^2_U)_{RR}$, $(\mathcal{M}^2_U)_{LR}$ are show here
\begin{eqnarray}
(\mathcal{M}^2_U)_{LL}&=&-\frac{e^2(v_1^2-v_2^2)(1-4c_W^2)}{24s_W^2c_W^2}+\frac{v_2^2Y_u^2}{2} +(Km_Q^2K^\dag)^T+\frac{g_B^2}{6}(v_B^2-\bar{v}^2_B),
\nonumber\\(\mathcal{M}^2_U)_{RR}&=&\frac{e^2(v_1^2-v_2^2)}{6c_W^2}+\frac{v_2^2Y_u^2}{2} +m_U^2-\frac{g_B^2}{6}(v_B^2-\bar{v}^2_B),
\nonumber\\(\mathcal{M}^2_U)_{LR}&=&-\frac{1}{\sqrt{2}}\Big(v_1(A_u'+Y_u\mu^*)+v_2A_u\Big).
\end{eqnarray}

6. Charginos mass matrix in BLMSSM is expressed as follow, and that existed in MSSM,
\begin{eqnarray}
Z_-^T
\left(\begin{array}{cc} M_2 & \frac{ev_2}{\sqrt{2}s_W} \\
                        \frac{ev_1}{\sqrt{2}s_W} & \mu
\end{array}\right)
Z_+=
\left(\begin{array}{cc}M_{\chi_1^\pm} & 0 \\
                          0 & M_{\chi_2^\pm}
\end{array}\right).
\end{eqnarray}

7. In BLMSSM, there are new exotic quarks $b^\prime$,
\begin{eqnarray}
W_{b^\prime}^\dag
\left(\begin{array}{cc} -\frac{1}{\sqrt{2}}\lambda_Qv_B & -\frac{1}{\sqrt{2}}Y_{d_5}v_u \\
                        -\frac{1}{\sqrt{2}}Y_{d_4}v_d   &  \frac{1}{\sqrt{2}}\lambda_d\bar{v}_B
\end{array}\right)
U_{b^\prime}=
\left(\begin{array}{cc}M_{b_4} & 0 \\
                          0 & M_{b_5}
\end{array}\right).
\end{eqnarray}

8. In BLMSSM, there are new exotic scalar particles $X$,

\begin{eqnarray}
Z^{\dag}_X\left(\begin{array}{cc}
  |\mu_X|^2+SS &-\mu_X^*B_X^* \\
    -\mu_XB_X & |\mu_X|^2-SS\\
    \end{array}\right)  Z_X=\left(     \begin{array}{cc}
 m_{X1}^2 &0 \\
    0 & m_{X2}^2\\
    \end{array}\right),
   \end{eqnarray}
   \begin{eqnarray}
\left(     \begin{array}{c}
  X_1 \\  X_2\\
    \end{array}\right) =Z_X^{\dag}\left( \begin{array}{c}
  X \\  X'^*\\
    \end{array}\right).
   \end{eqnarray}

The $Z_X$ is defined as unitary matrix to rotate rotate superfield mass squared matrix to mass eigenstates. Meanwhile, the mass matrix satisfy the following relation
\begin{eqnarray}
SS=\frac{g_B^2}{2}(\frac{2}{3}+B_4)(v_B^2-\bar{v}_B^2).
\end{eqnarray}

9. In the basis $\tilde{t}^{'T}=(\tilde{Q}^{1}_{4}, \tilde{U}^{c*}_{4}, \tilde{Q}^{2c*}_{5}, \tilde{U}_{5})$, $\tilde{b}^{'T}=(\tilde{Q}^{2}_{4}, \tilde{D}^{c*}_{4}, \tilde{Q}^{1c*}_{5}, \tilde{D}^{*}_{5}).$ The concrete forms for the exotic scalar quarks mass squared matrix are shown here,
\begin{eqnarray}
\noindent(\mathcal {M})_{\tilde{b}^\prime}^2(11)&=&m^2_{\tilde{Q}_4}+\frac{1}{2}Y^2_{u_4}v_u^2+\frac{1}{2}Y^2_{d_4}v_d^2+\frac{1}{2}\lambda^2_Qv_B^2
                                                -(\frac{1}{2}-\frac{2}{3}s_W^2)m_Z^2\cos2\beta+\frac{B_4}{2}m_{Z_B}^2\cos2\beta_B,
\nonumber\\ (\mathcal {M})_{\tilde{b}^\prime}^2(22)&=&m^2_{\tilde{D}_4}+\frac{1}{2}Y^2_{d_4}v_d^2+\frac{1}{2}\lambda^2_d\bar{v}_B^2-\frac{1}{3}s_W^2m_Z^2\cos2\beta
                                         -\frac{B_4}{2}m_{Z_B}^2\cos2\beta_B,
\nonumber\\(\mathcal {M} 0_{\tilde{b}^\prime}^2(33)&=&m^2_{\tilde{Q}_5}+\frac{1}{2}Y^2_{u_5}v_d^2+\frac{1}{2}Y^2_{d_5}v_u^2+\frac{1}{2}\lambda^2_Qv_B^2
                                          -(\frac{1}{2}+\frac{1}{3}s_W^2)m_Z^2\cos2\beta-\frac{1+B_4}{2}m_{Z_B}^2\cos2\beta_B,
\nonumber\\(\mathcal {M})_{\tilde{b}^\prime}^2(44)&=&m^2_{\tilde{D}_5}+\frac{1}{2}Y^2_{d_5}v_u^2+\frac{1}{2}\lambda^2_d\bar{v}_B^2
                                         +\frac{1}{3}s_W^2m_Z^2\cos2\beta+\frac{1+B_4}{2}m_{Z_B}^2\cos2\beta_B,
\nonumber\\(\mathcal {M})_{\tilde{b}^\prime}^2(21)&=&-\frac{1}{\sqrt{2}}Y_{d_4}v_dA_{d_4}+\frac{1}{\sqrt{2}}Y_{d_4}\mu v_d,
\nonumber\\(\mathcal {M})_{\tilde{b}^\prime}^2(31)&=&-\frac{1}{\sqrt{2}}\lambda_Qv_BA_{BQ}+\sqrt{2}\lambda_Q\mu_B\bar{v}_B,
\nonumber\\(\mathcal {M})_{\tilde{b}^\prime}^2(41)&=&-\frac{1}{\sqrt{2}}Y_{d_4}\lambda_dv_d\bar{v}_B+\frac{1}{\sqrt{2}}Y_{d_5}\lambda_Qv_uv_B,
\nonumber\\(\mathcal {M})_{\tilde{b}^\prime}^2(32)&=& \frac{1}{2}\lambda_QY_{d_4}v_dv_B+\frac{1}{2}\lambda_dY_{d_5}v_u\bar{v}_B,
\nonumber\\(\mathcal {M})_{\tilde{b}^\prime}^2(42)&=&-\frac{1}{\sqrt{2}}\lambda_dA_{BD}\bar{v}_B+\frac{1}{\sqrt{2}}\lambda_d\mu_Bv_B,
\nonumber\\(\mathcal {M})_{\tilde{b}^\prime}^2(34)&=&-\frac{1}{\sqrt{2}}Y_{d_5}A_{d_5}v_u+\frac{1}{\sqrt{2}}Y_{d_5}\mu v_d.
\end{eqnarray}

\section{NUMERICAL RESULTS\label{sec4}}
 As many supersymmetric extension, BLMSSM enlarged the gauge groups, and many parameters are introduced. The expression of mass matrices with these parameter are usually complicated.
 So it is difficult to obtain the possible parameter space with mass limits suggested by LHC observation. For example, $v_{bt}$, $tan\beta_B$, $g_B$ simultaneously appear in baryon neutralinos, down-squarks and up-squarks mass matrix. In addition, the baryon neutralinos mass matrix only the first diagonal elements is nonzero, so it's hard to pick up the appropriate value of $g_B$, $v_B$, $\bar{v}_B$ through $v_{bt}$, $m_{Z_{B}}$ in the process of diagonalization, makesure nonzero mass value of particle and meet mass limit.

In this section, we give the contour plot of mass vary with model parameters. By scanning the parameter space, we calculate the mass matrices numerically, then diagonalize them to get the masses of different particles.
In our numerical analysis, we adopt the following parameters in Table \ref{tab1}.

\begin{table}[!htbp]
\begin{tabular}
{@{}c@{  }c@{   }c@{   }c@{   }|c@{   }c@{   }c@{   }c@{}}\hline
$\alpha$ & 1/128 & $M_u$ & 0.0023 & $ A_{d_{4}}$ & 100 & $m^{2}_{D_{4}}$ & 2500\\
$\tan\beta$ & 5 & $M_d$ & 0.0048 & $ A_{d_{5}}$ & 100 & $m^{2}_{D_{5}}$ & 2500\\
$\lambda_Q$ & 0.5 & $M_c$ & 1.275 & $\mu_{X}$ & 1500 & $\mu$ & -1000\\
$\lambda_u$ & 0.5 & $M_s$ & 0.095 & $m_1$ & 1200 & $m_2$ & 1200\\
$\lambda_d$ & 0.5 & $M_t$ & 173.5 & $B_X$ & 400 & $m_{Z_B}$ & 1000\\
$\lambda_1$ & 0.4 & $M_b$ & 4.18 & $\tan\beta_B$ & 2.5 & $v_{bt}$ & 5000\\
$\lambda_3$ & 0.4 & $M_W$ & 80.385 & $\mu_{B}$ & 1100 & $m_B$ & 2000\\
$ A_{BQ}$ & 100 & $M_Z$ & 91.188 & $ A_{u_{4}}$ & 100 & $m^{2}_{U_{4}}$ & 2500\\
$ A_{BU}$ & 100 & $m^{2}_{Q_{4}}$ & 2500 & $ A_{u_{5}}$ & 100 & $m^{2}_{U_{5}}$ & 2500\\
$ A_{BD}$ & 100 & $m^{2}_{Q_{5}}$ & 2500 & & & &\\
\hline
\end{tabular}
\caption{parameters in the BLMSSM}
\label{tab1}
\end{table}

\begin{figure}[h]
\setlength{\unitlength}{1mm}
\centering
\begin{minipage}[c]{0.45\textwidth}
\includegraphics[width=2.9in]{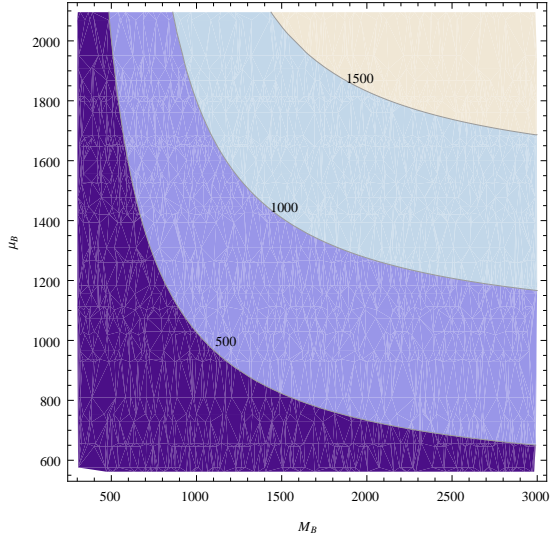}
\end{minipage}%
\caption[]{The contour for the mass of baryon neutralinos with respect to $M_B$ versus $\mu_B$ with the parameters are same as above in table 1.}
\label{fig1}
\end{figure}

The mass matrix of the baryon neutralinos includes $M_B$ and $\mu_B$.
$\mu_B$ is a Non-diagonal element and $M_B$ is a diagonal element of this mass matrix.
Therefore, the two parameters $M_B$ and $\mu_B$ can affect the contributions for the particle mass in some ways.
We assume the value of parameters are same as above in Table 1.
In Fig.\ref{fig1}, we show the contour for the mass of baryon neutralinos with respect to $M_B$ versus $\mu_B$.
With the increase of $M_B$ and $\mu_B$, the mass value of the baryon neutralinos also increase.
Here $M_B$ changes between 300 GeV and 3000 GeV and $\mu_B$ changes between 500 GeV and 2100 GeV.

\begin{figure}[h]
\setlength{\unitlength}{1mm}
\centering
\begin{minipage}[c]{0.45\textwidth}
\includegraphics[width=2.9in]{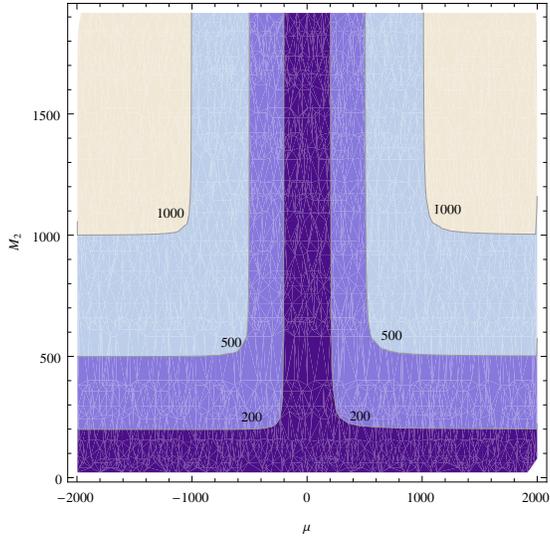}
\end{minipage}%
\caption[]{The contour for the mass of charginos or neutralinos with respect to $M_2$ versus $\mu$ with $M_B=1100$ GeV.}
\label{fig2}
\end{figure}

 $\mu$ and $M_2$ are not only the diagonal elements of the charginos mass matrix,
 but also constitute the mass matrix of neutralinos.
 Considerable influence to masses of charginos and neutralinos from $\mu$ and $M_2$ are hopeful.
 To see how $\mu$ and $M_2$ affect the numerical results, with $ M_B=1100 $ GeV.
 We give out the allowed region in the plane of $M_2$ versus $\mu$.
 Figure \ref{fig2} implies that when $\mu$ is near 0, the results are less than 200 GeV.
 The effects from $M_2$ are very weak, and can be neglected. The value of $\mu$ can vary from -2000 to 2000 GeV.
 The figure trend of the contour for the mass of neutralinos with respect to $M_2$ versus $\mu$ with $M_B=1100$ GeV is same as the Fig.\ref{fig2}.

\begin{figure}[h]
\setlength{\unitlength}{1mm}
\centering
\begin{minipage}[c]{0.45\textwidth}
\includegraphics[width=2.9in]{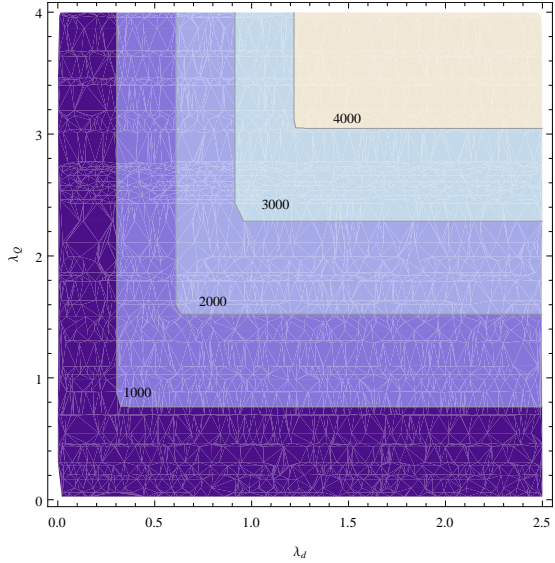}
\end{minipage}%
\caption[]{The contour for the mass of exotic -1/3 quark or exotic -1/3 squark with respect to $\lambda_{Q}$ versus $\lambda_{d}$ with $M_B=1100$ GeV.}
\label{fig3}
\end{figure}

 $\lambda_{Q}$ and $\lambda_{d}$ are not only the diagonal elements of the exotic -1/3 quark mass matrix,
 but also constitute the mass matrix of exotic -1/3 squark. Here we consider the elements of $\lambda_{Q}$ and $\lambda_{d}$,
 and suppose $ M_B=1100$ GeV. After the numerical calculation,
 the contour plot of $\lambda_{Q}$ versus $\lambda_{d}$ about the mass of exotic +2/3 quark and exotic -1/3 quark is shown in Fig.\ref{fig3}.
 The mass of these particles will get values less then 1000 GeV when decrease the value of $\lambda_{Q}$ and $\lambda_{d}$.
 At this point, the exotic -1/3 quark and the exotic -1/3 squark is likely to be found in LHC.

\begin{figure}[h]
\setlength{\unitlength}{1mm}
\centering
\begin{minipage}[c]{0.45\textwidth}
\includegraphics[width=2.9in]{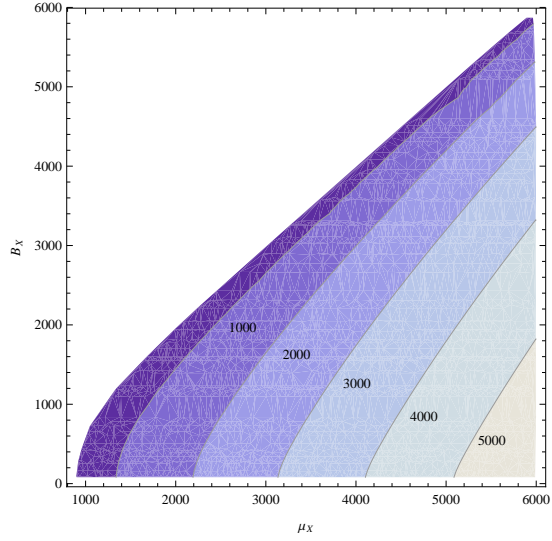}
\end{minipage}%
\caption[]{The contour for the mass of Superfields with respect to $\mu_{X}$ versus $B_{X}$
with $A_{BQ}=A_{BU}=A_{BD}=A_{u_{4}}=A_{u_{5}}=A_{d_{4}}=A_{d_{5}}=1000$  GeV, $\lambda_{Q}=0.7$ and $\tan\beta=3$.}
\label{fig4}
\end{figure}

In the figure \ref{fig4}, we plot the values of the $\mu_{X}$ and $B_{X}$ that lead to the mass value of Superfields.
We use $ A_{BQ}=A_{BU}=A_{BD}=A_{u_{4}}=A_{u_{5}}=A_{d_{4}}=A_{d_{5}}=1000$ GeV, $\lambda_{Q}=0.7$ and $\tan\beta=3$ for these plots.
Figure \ref{fig4} implies that from 0 to 6000  GeV of $B_{X}$ and from 1000 to 6000  GeV of $\mu_{X}$,
the mass values of superfields are all increasing functions of the enlarging $\mu_{X}$ and the diminishing $B_{X}$.
Moreover, there exist the value space that less then 1000 GeV, so this particle is likely to be found in LHC.
In addition, there has an large empty area, so we choose one point to analyze. After the numerical analysis, as $B_{X}=5000$ GeV,
$\mu_{X}=2000$  GeV, can lead to negative mass values of the superfields, so ones are unsuitable.

\begin{figure}[h]
\setlength{\unitlength}{1mm}
\centering
\begin{minipage}[c]{0.45\textwidth}
\includegraphics[width=2.9in]{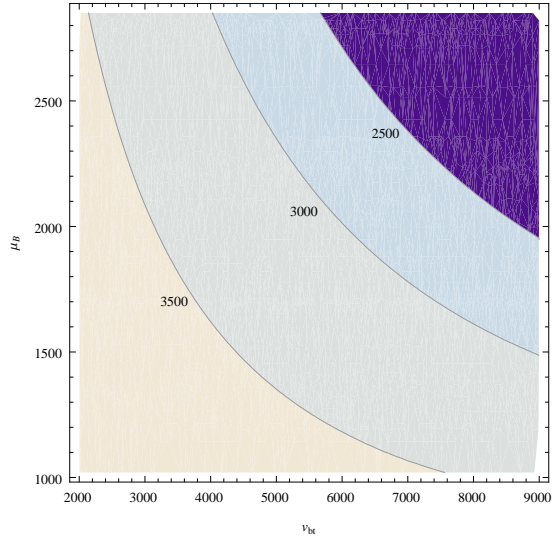}
\end{minipage}%
\caption[]{The contour for the mass of the exotic squarks with respect to $V_{bt}$ versus $\mu_{B}$
with $A_{BQ}=A_{BU}=A_{BD}=A_{u_{4}}=A_{u_{5}}=A_{d_{4}}=A_{d_{5}}=200$ GeV, $m^{2}_{Q_{4}} = m^{2}_{Q_{5}} = m^{2}_{U_{4}} = m^{2}_{U_{5}} = m^{2}_{D_{4}} = m^{2}_{D_{5}} = 4000$ GeV
and $\mu_{X}=2000 $ GeV.}
\label{fig5}
\end{figure}

Compared with MSSM, $V_{bt}$ and $\mu_{B}$ are new parameters that have relation with mass matrices of the exotic squarks $\tilde{b'}$ and $\tilde{t'}$.
Therefore, the effects to particles masses from $V_{bt}$ and $\mu_{B}$ are of interest.
In the plane of $V_{bt}$ versus $\mu_{B}$,
with $A_{BQ}=A_{BU}=A_{BD}=A_{u_{4}}=A_{u_{5}}=A_{d_{4}}=A_{d_{5}}=200$GeV, $m^{2}_{Q_{4}} = m^{2}_{Q_{5}} = m^{2}_{U_{4}} = m^{2}_{U_{5}} = m^{2}_{D_{4}} = m^{2}_{D_{5}} = 4000$ GeV
and $\mu_{X}=2000 $  GeV,
we show the allowed results denoted by the dots in Fig.\ref{fig5}.
The values of the exotic squarks are less then 2500 GeV when $V_{bt}$ over the value of 5400 and $\mu_{B}$ over the value of 2000  GeV.

\begin{figure}[h]
\setlength{\unitlength}{1mm}
\centering
\begin{minipage}[c]{0.45\textwidth}
\includegraphics[width=2.9in]{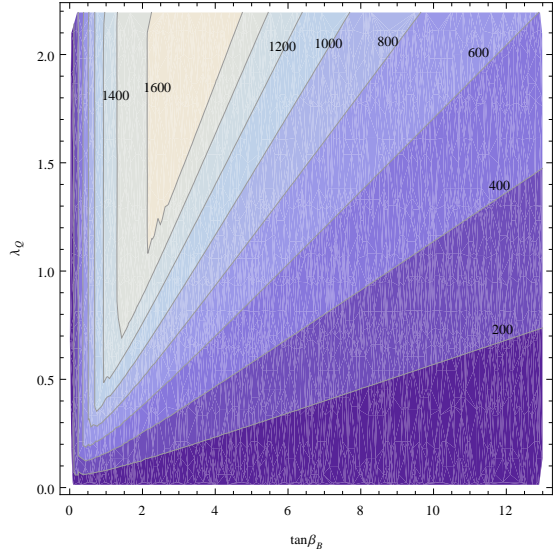}
\end{minipage}%
\caption[]{The contour for the mass of the exotic quarks with respect to
$\lambda_Q$ versus $\tan\beta_B$ with $\tan\beta=5, A_{BQ}=A_{BU}=A_{BD}=A_{u_{4}}=A_{u_{5}}=A_{d_{4}}=A_{d_{5}}=600$ and $M_B=1100$ GeV.}
\label{fig6}
\end{figure}

$\lambda_Q$ and $\tan\beta_B$ are important for the mass of the exotic quarks.
Therefore, the numerical results maybe influenced obviously by varying $\lambda_Q$ and $\tan\beta_B$.
For simplicity, we adopt $\tan\beta=5, A_{BQ}=A_{BU}=A_{BD}=A_{u_{4}}=A_{u_{5}}=A_{d_{4}}=A_{d_{5}}=600$ GeV and $M_B=1100$ GeV,
and plot $\lambda_Q$ varying with $\tan\beta_B$ in Fig.\ref{fig6} which $\lambda_Q$ changes between $0\sim2.2$ and $\tan\beta_B$ changes between $0\sim13$.
It implies that in the region$(0\sim1)$ of $\tan\beta_B$ the effect of $\lambda_Q$ are small, but in the region$(1\sim13)$ of $\tan\beta_B$ the effect of $\lambda_Q$ are strong,
when $\tan\beta_B$ takes a certain value, the mass value of exotic quarks is decreasing with the decreasing value of $\lambda_Q$.
Most masses of the particles are smaller than 1000 GeV.

\begin{figure}[h]
\setlength{\unitlength}{1mm}
\centering
\begin{minipage}[c]{0.45\textwidth}
\includegraphics[width=2.9in]{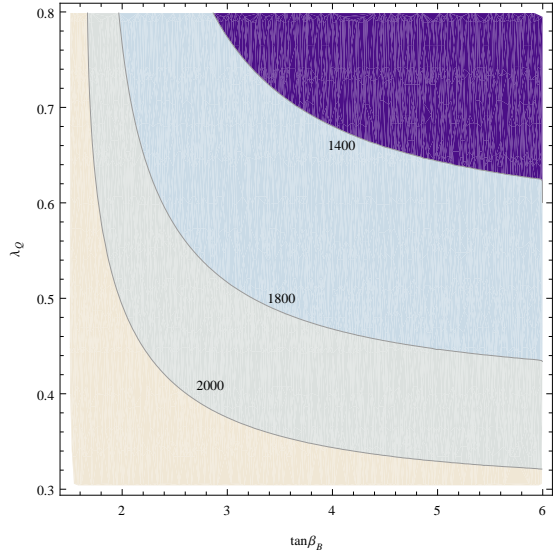}
\end{minipage}%
\caption[]{The contour for the mass of the exotic squarks with respect to $\lambda_Q$ versus $\tan\beta_B$
with $\tan\beta=5, A_{BQ}=A_{BU}=A_{BD}=A_{u_{4}}=A_{u_{5}}=A_{d_{4}}=A_{d_{5}}=600$ GeV and $M_B=1100$ GeV.}
\label{fig7}
\end{figure}

Now, let us investigate the results for the mass of the exotic squarks where one takes into account the $\lambda_Q$ and $\tan\beta_B$.
In order to illustrate the numerical solutions we choose $\tan\beta=5, A_{BQ}=A_{BU}=A_{BD}=A_{u_{4}}=A_{u_{5}}=A_{d_{4}}=A_{d_{5}}=600$ GeV and $M_B=1100$ GeV.
In Fig.\ref{fig7}, we show the values for the mass of the exotic squarks when $\lambda_Q$ changes between $0.3\sim0.8$ and $\tan\beta_B$ changes between $1.5\sim6$.
The mass values of the exotic squarks decrease when increase the values of $\lambda_Q$ and $\tan\beta_B$, simultaneously.

\begin{figure}[h]
\setlength{\unitlength}{1mm}
\centering
\begin{minipage}[c]{0.45\textwidth}
\includegraphics[width=2.9in]{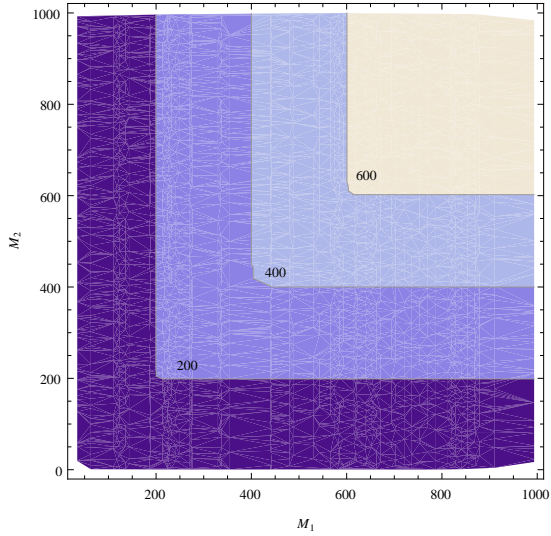}
\end{minipage}%
\caption[]{The contour for the mass of the exotic squarks with respect to $M_2$ versus $M_1$ with $M_B=1100$ GeV.}
\label{fig8}
\end{figure}

$M_2$ and $M_1$ are diagonal elements of neutralinos mass matrix.
They are sensitive parameters and affect the mass of neutralinos forcefully. Here we use the parameter $M_B=1100$ GeV, in the plane of $M_2$ versus $M_1$,
the contour plot is scanned, and the allowed results are shown in Fig.\ref{fig8}. When $M_2$ and $M_1$ are greater than 600 GeV, the mass value of neutralinos is larger then 600 GeV.
On the contrary, When $M_2$ and $M_1$ are less than 200 GeV, the mass value of neutralinos is smaller then 200 GeV.
These are an acceptable kinematic range for discovery at the LHC.

\begin{figure}[h]
\setlength{\unitlength}{1mm}
\centering
\begin{minipage}[c]{0.45\textwidth}
\includegraphics[width=2.9in]{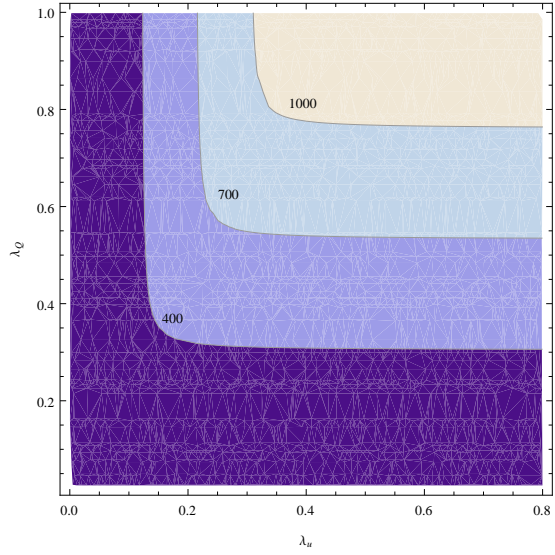}
\end{minipage}%
\caption[]{The contour for the mass of the exotic quarks with respect to $\lambda_Q$ versus $\lambda_u$ with the values of parameters are same as above in table 1.}
\label{fig9}
\end{figure}

$\lambda_Q$ is related to the mass matrices of exotic quarks.
Through scanning the contour plot of $\lambda_Q$ versus $\lambda_u$ in Fig.\ref{fig9} with the values of parameters are same as above in table 1,
we find that the most mass values of exotic quarks are less then 1000 GeV, when $\lambda_Q$ in the region $0\sim1$ and $\lambda_u$ in the region $0\sim0.8$.
Considering the experiment about LHC, one can generate this particle.

\begin{figure}[h]
\setlength{\unitlength}{1mm}
\centering
\begin{minipage}[c]{0.45\textwidth}
\includegraphics[width=2.9in]{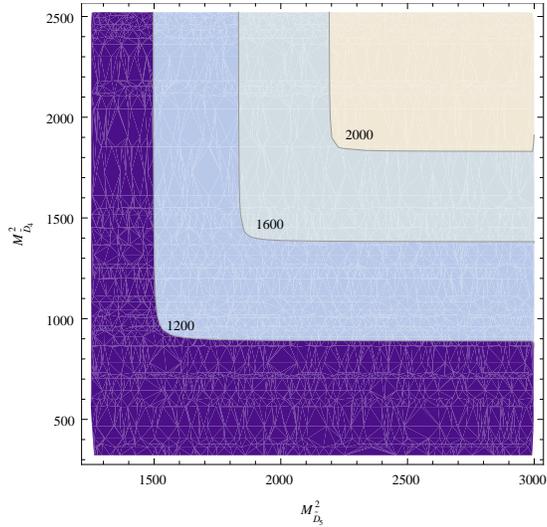}
\end{minipage}%
\caption[]{The contour for the mass of the exotic -1/3 squark with respect to $m^2_{\tilde{D}_{4}}$ versus $m^2_{\tilde{D}_{5}}$ with $\lambda_Q=0.1$, $\lambda_d=0.1$.}
\label{fig10}
\end{figure}

 Here, we consider $m^2_{\tilde{D}_{4}}$ versus $m^2_{\tilde{D}_{5}}$, ones are diagonal elements that included in the mass matrice of the exotic -1/3 squark,
 which should produce considerable influence on the numerical results. Based on the supposition $\lambda_Q=0.1$, $\lambda_d=0.1$,
 we scan the contour of $m^2_{\tilde{D}_{4}}$ versus $m^2_{\tilde{D}_{5}}$ in Fig.\ref{fig10}. From 300 to 2600 of $m^2_{\tilde{D}_{4}}$,
 the mass values of the exotic -1/3 squark are all increasing functions of the enlarging $m^2_{\tilde{D}_{5}}$. The values of $m^2_{\tilde{D}_{5}}$ vary from 1250 to 3000.
 The values of $m^2_{\tilde{D}_{4}}$ vary from 300 to 2600 GeV.

\begin{figure}[h]
\setlength{\unitlength}{1mm}
\centering
\begin{minipage}[c]{0.45\textwidth}
\includegraphics[width=2.9in]{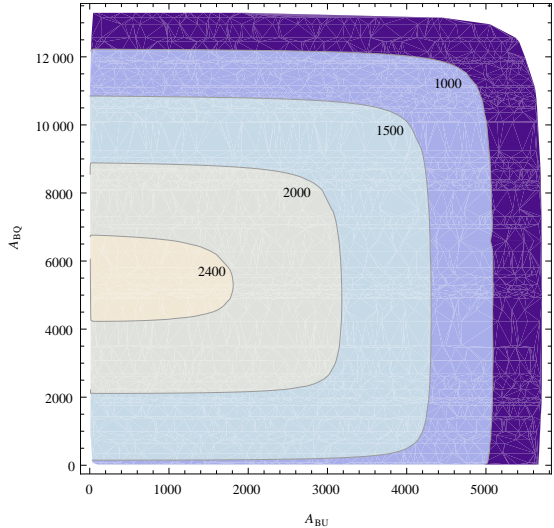}
\end{minipage}%
\caption[]{The contour for the mass of the exotic +2/3 squark with respect to $ A_{BU}$ versus $ A_{BQ}$ with $A_{BD}=A_{u_{4}}=A_{u_{5}}=A_{d_{4}}=A_{d_{5}}=6000$ GeV, $M_B=1800$ GeV and $\lambda_Q=0.7$.}
\label{fig11}
\end{figure}

In fig.\ref{fig11} we show the results of the $ A_{BU}$ and $ A_{BQ}$ values by changing the mass of the exotic +2/3 squark
and assuming $A_{BD}=A_{u_{4}}=A_{u_{5}}=A_{d_{4}}=A_{d_{5}}=6000$ GeV, $M_B=1800 GeV$ and $\lambda_Q=0.7$.
In this way, we can achieve a exotic +2/3 squark mass less then 1000 GeV for the region when $ A_{BU}$ is almost over 5000 GeV  and $ A_{BQ}$ is almost over 12000 GeV.

\section{CONCLUSITION\label{sec5}}
In this work, we study the relation between parameters of BLMSSM model and the masses of supersymmetric particles.
With the model introduced firstly, we collected the mass matrices.
To show the impact on masses, parameters such as $M_B, \mu_B$ contained in baryon neutralions,
couplings of exotic quarks and new Higgs field $\lambda_Q,\lambda_d$, the $\tan_{\beta_B}$ are scanned to give the masses numerically
by diagonalizing the mass matrices. From the contour plots, we can intuitively obtain the parameter space for mass limits of baryon neutrilinos, charginos,
exotic quarks and squarks, and so on.

\begin{acknowledgments}
\indent\indent
The work has been Supported by National Natural Science
Foundation of China (Grant No. 11347185) and Scientific and Technologial Innovation Programs of Higher Education Institutions in Shanxi(Grant No.2017113)
\end{acknowledgments}


\begin{thebibliography}{99}
\bibitem{bibitem1} F. Gianotti, CERN Public Seminar, Update on the Standard Model Higgs searches
in ATLAS, December 2011.

\bibitem{bibitem2} G. Tonelli, CERN Public Seminar, Update on the Standard Model Higgs searches
in CMS, December 2011.

\bibitem{bibitem3} J. Rosiek, Phys.Rev. D {\bf41} (1990) 3464; arXiv:hep-ph/9511250;

\bibitem{bibitem4} Tai-Fu Feng, Xiu-Yi Yang, Nucl.Phys. B {\bf814} (2009) 101;

\bibitem{bibitem5} H.P. Nilles, Phys. Rept. {\bf110} (1984) 1;

\bibitem{bibitem6} H.E. Haber and G.L. Kane, Phys. Rept. {\bf117} (1985).

\bibitem{bibitem7} P.F. Perez and M.B. Wise, J. High Energy Phys. {\bf08} (2011) 068.

\bibitem{bibitem8}  P.F. Perez, Phys. Lett. B {\bf711} (2012) 353.

\bibitem{bibitem9}  J.M. Arnold, P.F. Perez, B. Fornal, and S. Spinner, Phys. Rev. D {\bf85} (2012). 115024.

\bibitem{bibitem10} N. Escudero, D.E.L. Fogliani, C. Munoz, et al., J. High Energy Phys. {\bf12} (2008) 099.

\bibitem{bibitem11}P. Ghosh and S. Roy, J. High Energy Phys. {\bf04} (2009) 069.

\bibitem{bibitem12}   A. Bartl, M. Hirsch, S. Liebler, et al., J.High Energy Phys. {\bf05} (2009) 120.

\bibitem{bibitem13} P. Fileviez Perez and S. Spinner, Phys. Lett. B{\bf673}(2009)251.

\bibitem{bibitem14} V. Barger, P. Fileviez Perez, and S. Spinner, Phys. Rev. Lett.{\bf102}(2009)181802.

\bibitem{bibitem15} P. Fileviez Perez and S. Spinner, Phys. Rev. D{\bf80}(2009)015004.

\bibitem{bibitem16} P. Fileviez Perez and S. Spinner, JHEP {\bf1204}(2012)118.

\bibitem{bibitem17} P. Fileviez Perez and S. Spinner, Phys. Rev. D{\bf83}(2011)035004.

\bibitem{bibitem18} P.F. Perez, Phys. Lett. B {\bf711} (2012) 353.

\bibitem{bibitem19} J.M. Arnold, P.F. Perez, B. Fornal, S. Spinner, Phys. Rev. D {\bf85} (2012) 115024.

\bibitem{bibitem20} R. Barbieri and A.Masiero, Nucl. Phys. B{\bf267}, 679 (1986).

\bibitem{bibitem21}  S.Dimopoulos and L. J. Hall, Phys. Lett. B {\bf207}, 210 (1988).

\bibitem{bibitem22} For a review see: P. Nath and P. Fileviez Perez, Phys. Rept. {\bf441}.
(2007) 191 [arXiv:hep-ph/0601023].

\bibitem{bibitem23} S. M. Zhao, T. F. Feng, B. Yan et al, JHEP, {\bf10}: 020(2013)

\bibitem{bibitem24} S. M. Zhao, T. F. Feng, H. B. Zhang et al, JHEP,{\bf11}: 119 (2014)

\bibitem{bibitem25} F. Sun, T. F. Feng, S. M. Zhao, et al, Nucl. Phys.B, {\bf888}: 30-51 (2014)

\bibitem{bibitem26} P.F. Perez and M.B. Wise, Phys. Rev. D {\bf82} (2010) 011901.

\bibitem{bibitem27} Tai-Fu Feng, Shu-Min Zhao, Hai-Bin Zhang, Yin-Jie Zhang, Yu-Li Yan, Nucl.Phys. B {\bf871}
(2013) 223.

\bibitem{bibitem28} Xing-Xing Dong, Shu-Min Zhao, Hai-Bin Zhang, et al., Chin. Phys. C {\bf40} (2016) 093103.

\bibitem{bibitem29} S. M. Zhao, T. F. Feng, X. J. Zhan et al, JHEP, {\bf07}: 124 (2015)

\bibitem{bibitem30} S. M. Zhao, T. F. Feng, H. B. Zhang et al, Phys.
Rev. D, {\bf92}: 115016 (2015)
\end{thebibliography}
\end{document}